\documentstyle[12pt]{article}
\newcommand{\Dt}{\Delta t}
\newcommand{\lc}{\lambda_c}
\newcommand{\Dx}{\Delta x}
\newcommand{\LL}{\tilde {\nabla}^2}
\newcommand{\rb}{$\overline {\rho}$}

\begin{document}
\title{Numerical Study of a Field Theory for Directed Percolation}
\author{Ronald Dickman\\
        Department of Physics and Astronomy\\
        Lehman College, CUNY\\
        Bronx, NY 10468-1589\\}
\date{}
\maketitle
\begin{abstract}
A numerical method is devised for study of stochastic
partial differential equations (SPDEs)
describing directed percolation,
the contact process, and other
models with a continuous transition to an absorbing state.
Owing to the heightened sensitivity to fluctuations
attending multiplicative noise in the vicinity of an
absorbing state,
a useful method requires discretization of
the field variable as well as of space
and time.  When applied to the field theory for directed percolation
in 1+1 dimensions, the method yields critical
exponents which
compare well against accepted values.
\end{abstract}
\vspace{2em}
\begin{centering}
\end{centering}
\\

\newpage
\section{Introduction}

The study of critical phenomena in simple nonequilibrium lattice
models has reached the stage where many transitions can be
assigned to one of a small set of universality classes.   For
continuous transitions into an absorbing state, a very high degree of
universality
has been found, with many examples supporting the prediction
\cite{janssen1,pgrg,geoff}
that such transitions belong generically to the class of directed
percolation (DP).
Examples include the basic contact process and its variants
\cite{harris,liggett,durrett,tdpdif,tdpcpgen},
surface reaction models \cite{zgb,geoff,iwanzgb},
branching and annihilating random walks with
odd parity \cite{taktre,iwanbaw,baw4ann},
assorted multiparticle processes \cite{aukrust,tripann,tripcrea,park},
and even models with multiple
absorbing configurations \cite{pcppre,albano,ditri,iwancono,mendes}.
Each is an interacting particle system characterized
by rules for elementary processes such as
creation, annihilation, and diffusion.
Looking at the rules, there is little to tell us what sort of critical
behavior to expect, nor why it is universal.  Understanding
of universality emerges
instead from the study of coarse-grained
formulations which capture the large-scale features essential
to critical behavior.  In such field theories the microscopic picture of
particles on
a lattice is replaced by a set of densities which evolve {\em via}
stochastic partial differential equations (SPDEs).
At this level,
renormalization group methods may be applied.
\cite{janssen1,janssen2,grasund,elderschlogl,elderwilby}.
A basis for universality appears if one can show that the
continuum descriptions for various models differ only by irrelevant terms.
At present, however, there are many more models known (on
the basis of numerical work - simulation and/or series analysis),
to have DP critical behavior, than have been studied using field theory.
Useful continuum descriptions of multiparticle processes, for example,
have yet to be devised \cite{pgrg,pgkink}.

It is of interest, therefore, to study SPDEs
for nonequilibrium systems, and to compare
their behavior with the lattice models they are supposed to
describe.  But solving a nonlinear SPDE is not generally
feasible by analytic means, and so numerical methods
must be sought \cite{seepet}.  Numerical integration has been
applied to several SPDEs, for example
the time-dependent Ginzburg-Landau equation describing
phase separation \cite{rogers,gawl,ram}, and the Kardar-Parisi-Zhang
equation \cite{guo,chakra,kpzrefs}.
In problems with an absorbing state, however,
the usual approach does not yield useful results.
A method for dealing with such systems is proposed in
the present work, and is used to study the
field theory for the contact process.

The outline is as follows. In Section 2 I describe the
original model and the corresponding SPDE.
The integration scheme is introduced in Section 3, and
results are presented in Section 4. A discussion
and summary follow in Section 5.
\vspace{1em}

\section{Lattice Model and Field Theory}

In the {\em contact process}
(CP) \cite{harris},
each site of  the d-dimensional cubic lattice,
{\bf Z}$^{d}$, is either vacant, or is occupied
by a particle.  The transition rules are easily stated: a
vacant site with $n$
occupied nearest neighbors becomes occupied
at rate $\lambda n/2d $, and particles disappear at unit rate,
independent of their surroundings.
Evidently the vacuum is absorbing; the active phase,
characterized by a nonzero stationary particle density, \rb,
exists only for sufficiently large $\lambda$
(and only, strictly speaking, in the infinite-volume limit).
There is a continuous transition from the vacuum to the
active phase at a critical value $\lc$ \cite{bezu}.  (In one dimension
$\lc \simeq 3.2978$ \cite{tdpjsp}.)  The transition belongs to the
universality class of directed percolation.  (Note that the d-dimensional
CP corresponds to directed percolation in d+1 dimensions.)

Janssen \cite{janssen1} proposed a continuum description of the CP
and allied models:
\begin{equation}
\label{field}
\frac {\partial \rho ({\bf x},t)}{\partial t} = a \rho ({\bf x},t)
- b \rho ^{2} - c \rho ^{3} + \cdots  + D \nabla^{2} \rho + \eta ({\bf x},t).
\end{equation}

\noindent $\rho({\bf x},t) \geq 0$ is the coarse-grained particle
density; the ellipsis represents terms of higher order in $\rho$.
$\eta ({\bf x},t)$ is
a Gaussian noise, which respects the absorbing state ($\rho = 0$)
by virtue of the covariance:

\begin{equation}
\overline {\eta({\bf x},t) \eta({\bf x'},t')} \propto
\rho ({\bf x},t) \delta({\bf x-x'}) \delta (t-t').
\label{cov}
\end{equation}

\noindent This form can be justified by
coarse graining the CP, in the limit of large
bin size.  Let $n_{\rm i}$ be the number of particles
in bin i, and $\Delta n_{\rm i} $ the change in this number during a brief
interval.  The latter has expectation $\overline {\Delta n_{\rm i}}
\propto a n_{\rm i} + {\cal O}(n^2 _{\rm i})$,
(with $a \propto \lambda - 1 $), and
under the customary assumption of Poissonian
statistics for reaction systems,
its variance equals $\overline {\Delta n_{\rm i}}$.
For sufficiently large bins
we may approximate $\Delta n_{\rm i} $by a Gaussian.
Thus, since reactions in different bins are uncorrelated,
coarse-graining the original model leads to a
stochastic field theory with Gaussian noise whose
autocorrelation is proportional to the local density.
(There is also noise due to the
fluctuating diffusive current.  But
diffusive noise does not affect
the critical behavior in the present case, and so I shall
ignore it, in the interest of simplicity.)
Since Eq (\ref{field}) involves multiplicative noise, one must
decide upon an interpretation \cite{gardinerbook}.
As shown in the following section, the
Ito interpretation of Eq (\ref{field}) is demanded by
physical considerations.

In mean-field approximation (the spatially uniform, noise-free
version of Eq (\ref{field})), the vacuum
becomes unstable when $a=0$, and for $ a, b > 0 $ there is
an active state.   When fluctuations are taken into account,
the critical point shifts to $a_c > 0$, and the critical behavior
is nonclassical.  For example, the stationary density in the CP
scales as \rb $\propto (a - a_c)^{\beta}$, with $\beta \simeq 0.277 $
in one dimension.  (In mean-field theory, $\beta = 1$.)
Field-theoretic analysis \cite{janssen1,grasund}
reveals that the cubic and higher-order terms are irrelevant
to critical behavior, so long as $b>0$.
(Such terms are therefore ignored in what follows.)
The situation is analogous
to that in equilibrium critical phenomena, where the Ising universality
class is generic for models with a scalar order parameter
and short-range interactions.

Without noise,  Eq (\ref{field}) is a
reaction-diffusion equation, which
exhibits a mean-field critical point.
It is perhaps surprising that driving a reaction-diffusion equation with
multiplicative noise leads to the proper
exponents.  Of course the condition expressed in Eq (\ref{cov}) is
crucial in this regard.  On the other hand, it is not
clear whether adding a properly scaled noise
to the reaction diffusion equation always
yields a useful field theory \cite{pgrg,pgkink}.

Further unanswered questions are whether solutions to
Eq (\ref{field}) exist, and if so, whether they reproduce the
phenomenology of the lattice models they are supposed to describe.
(For example, Can the field actually fluctuate into the vacuum?)
Such issues never arise in
renormalization group analyses, where
the SPDE merely serves as a basis for
perturbation theory, which proceeds by expanding the formal
solution.  Since the exponents emerging from the $\epsilon$-expansion
analysis of Eq (\ref{field}) are in good agreement with series and
simulation results, there is no reason to doubt its validity
in this context.  The present work is concerned with
nonperturbative (numerical) solutions to a
discretized version of the SPDE.

\section{Numerical Method}

Can Eq (\ref{field}) be integrated numerically?
To begin, we discretize space,
obtaining a set of
Langevin equations which in one dimension
take the form

\begin{equation}
\label{lle}
\frac {d \rho (i,t)}{d t} =  a \rho (i,t)
- b \rho ^{2}  + D \LL \rho + \eta (i,t),
\end{equation}
where $i$ is a site index, (for convenience we assume
a spacing $\Dx = 1$
in the discretization),
and $\LL \rho(i,t) \equiv \rho(i+1,t) + \rho(i-1,t)
-2 \rho(i,t) $ is the
lattice Laplacian operator.  The noise term satisfies
$\overline {\eta(i,t) \eta(j,t')} = \Gamma
\rho (i,t) \delta_{i,j} \delta (t-t').$
($\Gamma $ is related to the growth rate $\lambda$
in the CP.  We may regard it as constant over the
range of parameter values of interest here, and set
$\Gamma = 1$ from here on.)
Applying the Cauchy-Euler method to these equations \cite{gardinerbook},
we find

\begin{equation}
\label{euler}
\rho(i,t+\Dt)- \rho(i,t) = [ a \rho(i,t) - b \rho(i,t) ^{2} +D\LL \rho(i,t)]\Dt
  + \sqrt {\rho(i,t) \Dt} Y(i,t).
\end{equation}

\noindent where the $Y(i,t)$ are independent Gaussians with zero mean and unit
variance.  Eq (\ref{euler}) is similar to the set of stochastic
differential equations (SDEs) derived
in Ref.\cite{seepet} from discretization of a SPDE.
The latter scheme, however, is not useful here, due to
the dominance of the noise in the vicinity of the critical
point.  Indeed, once we discretize time
there is nothing to prevent $\rho (i,t) $ becoming negative.
We might attempt to remedy this by stipulating that whenever
integration yields $\rho(i,t) < 0$, the density at
that site be set to zero.  But this artifice is not without drawbacks.
We are interested in the critical region, where $\rho $ is small.  If
the typical magnitude of the first term on the r.h.s. is $\epsilon$,
the noise term is of order $\sqrt \epsilon $, and so
overwhelms the deterministic part of the evolution.
In the original equation, the cumulative effect of
the systematic term is not obliterated by the noise,
which has zero mean.
But this is no longer so if
we truncate the noise in an unsymmetric
manner.
In simulations using the $\max [\rho, 0]$
rule, the process never reaches the vacuum (even for $a < 0$).
Regions of density zero are rapidly repopulated
by nearby active sites.
It appears, then, that straightforward
numerical integration of Eq (\ref{field}) is not useful.

Consider, for the moment, the Ito SDE
(a Malthus-Verhulst
process), with the same local terms as Eq (\ref{field}):
\begin{equation}
d \rho = [a \rho - b \rho^2]dt + \sqrt {\rho} d\xi(t),
\label{onedeg}
\end{equation}
with $\overline {d\xi(t)^2} = dt.$
Eq (\ref{onedeg}) describes Brownian motion in a potential
which grows $\propto \rho^3 $ for large $\rho$, and
has a minimum (for $a$, $b > 0$), at $\rho = a/b$.  There is an absorbing
boundary at $\rho = 0$, corresponding to the vacuum
in the CP.  Now suppose we had interpreted Eqs (\ref{field})
and (\ref{onedeg}) as {\em Stratonovich} equations.  The Ito
SDE corresponding to the Stratonovich
interpretation of Eq (\ref{onedeg}) is \cite{gardinerbook}
\begin{equation}
d \rho = [a \rho - b \rho^2 + \frac{1}{4}]dt + \sqrt {\rho} d\xi(t).
\label{strat}
\end{equation}
It includes a {\em constant source term}, so that $\rho =0$
is no longer absorbing.  Clearly this is not the problem we
began with!  Hence Eqs (\ref{field})
and (\ref{onedeg}) should be taken in the Ito sense.

Integrating Eq(\ref{onedeg}),
we have:
\begin{equation}
\Delta \rho = [a \rho - b \rho^2] \Delta t + \sqrt{\rho} \Delta W,
\label{odint}
\end{equation}
where $\Delta W = \sqrt{\Delta t} Y$, and $Y$ is Gaussian
with zero mean and unit variance.
Now to prevent $\rho + \Delta \rho $
going negative, I propose to discretize the density by setting
$ \rho = n \rho_{min} $, ($n \geq 0$),
and at the same time to truncate $Y$ symmetrically
by restricting its magnitude so: $|Y| \leq  Y_{max}$.
We require $ Y_{max} \sqrt{\Delta t} \leq \sqrt{\rho_{min}}$
to avoid negative densities.
This can be achieved in a variety of ways, for example by setting
\begin{equation}
 Y_{max} = \frac{|\ln \Delta t|}{3},
\label{ymax}
\end{equation}
and
\begin{equation}
\rho_{min} = \frac {(\ln \Delta t)^2 \Delta t}{9},
\label{rmin}
\end{equation}

\noindent Eqs (\ref{ymax}) and (\ref{rmin}) represent but one of an
infinity of choices.
Determining which is optimal for a
specific problem is left as a subject for future work.
I take $\rho_{min} \propto \Dt$ in hopes of minimizing the
effect of a discretized density.
The relatively slow growth of $Y_{max} $ poses
no essential difficulty.
(Note that for $\Delta t = 10^{-4}$ we have $ Y_{max} =
3.07$ --- about three standard deviations.)
Indeed, all noise distributions having
zero mean and finite variance should
yield qualitatively similar behavior.  If one were
interested solely in universal properties,
the Gaussian could be replaced with a uniform
distribution, in the interest of computational efficiency.

Having discretized $\rho$, we can define an integer process by exploiting
the invariance of Eq (\ref{onedeg}) under the rescaling:

\begin{equation}
b \rightarrow b' = \alpha b,
\label{rscb}
\end{equation}
\begin{equation}
\rho \rightarrow \rho' = \frac{\rho}{\alpha},
\label{rscr}
\end{equation}
\begin{equation}
\xi \rightarrow \xi' = \frac {\xi}{\alpha},
\label{rsce}
\end{equation}

If we choose $ \alpha = \rho_{min}$, then $\rho'$ is restricted to
integers $\geq 0$.
Discretization (in time) of Eq (\ref{onedeg}) leads to a
noise term $Y \sqrt {\rho \Delta t}$;
in the rescaled equation it becomes $Y \sqrt {\rho' \Delta t/\alpha} $.
($Y$ is a zero-mean, unit-variance Gaussian, truncated as
described above.)

We now have a discretized version of
Eq (\ref{onedeg}), in which positivity and zero-mean
noise are ensured at the cost of a ``quantized" density.
Since $\rho'$ can change only by integer steps, it is
likely (especially for small $\rho'$), that many increments
of the density will be rejected, for being of less
than unit magnitude.
It therefore seems advisable to introduce a
continuous variable $\psi$ which accumulates the increments
in density at each time step.  Whenever $|\psi| \geq 1$, the
integer part is transferred to $\rho'$.

In summary, the numerical scheme for Eq (\ref{onedeg}) is as follows.
At each time step $\psi \rightarrow \psi + \Delta \psi$,
where
\begin{equation}
\Delta \psi = (a \rho' - b' \rho'^2) \Delta t
+ Y \sqrt{\rho' \Delta t /\alpha},
\label{psi}
\end{equation}
and
\begin{equation}
\rho' \rightarrow \rho' + [\psi],
\end{equation}
\begin{equation}
\psi \rightarrow \psi - [\psi],
\label{psi2}
\end{equation}
where square brackets denote the integer part.
(Initially,
$\psi $ is zero.)
Eqs (\ref{psi}) - (\ref{psi2}) may be
viewed as a Malthus-Verhulst
process in which the population
change, $\Delta n \equiv \Delta \rho'$, is approximated
by a suitably truncated Gaussian random variable.
Simulations show the population
fluctuating around a quasi-steady value, $ \rho_{qs} \approx a/b$,
and eventually becoming trapped at zero (see the inset of Fig 1).
In Fig 1 the mean density for a sample of $10^3$ trials is
plotted for several time increments.  (The model parameters are
$a = 1.5$, $b=1$; $\rho = 1.6$ initially.)
The density decays exponentially, with relaxation times
12.9, 9.7, and 9.8 for time increments $10^{-3}$, $2 \times 10^{-4}$,
and $10^{-4}$, respectively.  This is in good agreement with the mean
first passage time, 10.3, for hitting $\rho = 0$.
(The latter is obtained from the
Fokker-Planck equation corresponding to
Eq (\ref{onedeg}) \cite{gardinerbook}.)

Our treatment of the SPDE,
Eq (\ref{field}), closely parallels that
of Eq (\ref{onedeg}).
Discretization and rescaling of Eq(\ref{lle}) yields a
set of diffusively coupled
Malthus-Verhulst processes:
\begin{equation}
\Delta \psi_i = (a \rho'_i - b' \rho'^{2}_{i} +D\LL \rho'_i ) \Delta t
+ Y_i \sqrt{\rho'_i \Delta t /\alpha},
\label{lpsi}
\end{equation}
and
\begin{equation}
\rho'_i \rightarrow \rho'_i + [\psi_i],
\end{equation}
\begin{equation}
\psi_i \rightarrow \psi_i - [\psi_i].
\end{equation}

\noindent (The
rescaling of Eqs (\ref{rscb}) - (\ref{rsce}) does not affect
the diffusion coefficient.)
We have converted the original SPDE into a
lattice of discrete stochastic
processes which
approaches the continuum model as $\Dt$
and $\Dx \rightarrow 0$.

A final technical point is that when integrating the
coupled equations,
the three parts of the evolution --- deterministic on-site contributions,
the noise term, and diffusion --- are implemented separately,
in turn (over the entire lattice), at each step.
(Transfer from $\psi $ to $\rho '$ is made
following each of the three sub-steps.)
Unphysical events such as
a site becomming empty, and subsequently acting
as a source for its neighbor, are eliminated
by this measure.

\section{Results}

I applied the scheme detailed above to systems of
several hundred to several thousand sites, in order to
study the critical behavior of Eq (\ref{field}).
To begin, I describe results for the steady-state,
obtained in a series of long runs (of duration
$t_f \approx 10^{4}$), on lattices of 500 --- 2000
sites, using a time step $\Dt = 10^{-4}$. The coefficient
$b$ was set to unity; diffusion rates $D = 1$ and 10 were
considered.
The stationary density \rb (a) (expressed in its
original units, prior to rescaling), is shown
in Fig 2.   The data suggest a
continuous transition to the vacuum at a critical value $a_c$
($\simeq 0.77$, 0.36, for $D= 1$, and 10, respectively).
To estimate the order-parameter
exponent $\beta$, one must estimate the critical value $a_c$,
and then plot the density versus $\Delta = a - a_c$ on log scales.
For $D=1$, a reasonably linear plot (for small $\Delta$) is obtained
when we choose $a_c$ = 0.769 (see Fig 3).  A least-squares linear fit to the
five points nearest $a_c$ yields a slope $\beta = 0.295$, in fair
agreement with $\beta =\ 0.277$ for the one-dimensional CP.
As is often the case, the slope depends quite
sensitively upon one's estimate of the critical point,
and so this analysis is not very precise.
(Taking $a_c = 0.765$, for example, one finds $\beta \approx 0.39$.)
The slope ($\approx 0.45$) obtained from the $D=10$ data
suggests a more mean-field like behavior for faster diffusion.
Indeed,  $a_c$ appears to shift towards its
mean-field value, 0, with increasing $D$.  For larger $D$ a
crossover from mean-field to DP-like behavior
presumably occurs very near the critical point.

In order to derive quantitative results on critical
behavior, I turn to the ``time-dependent" method
\cite{pgrg,torre}, in which one studies the dynamics of
spreading from a distribution localized near $x=0$.
The quantities of interest are the survival
probability, $P(t)$, mean total density $n(t)$,
and mean-square spread $R^2(t)$, for a large sample
of independent trials, all with the same
initial condition.  $P(t)$ denotes the probability
of {\em not} being in the vacuum state at time t,
$n(t)$ is the sum of the site densities (averaged
over all trials, {\em including} those which have
reached the vacuum by time $t$), and
$R^2(t) \equiv \sum_j j^2 \rho(j,t) / n(t)$.
In a subcritical system, ($a < a_c$), we
expect $P$ and $n$ to decay exponentially, whilst
$R^2(t) \propto t$.  For $a > a_c$, $P$ approaches
a nonzero limiting value, $n(t) \simeq t^d$ (in $d$
dimensions), and $R^2(t) \simeq t^2$, as a fraction
of trials survive indefinitely and spread at finite
speed into the surrounding vacuum.  At the critical
point there is no characteristic time-scale
for relaxation, and the evolution is characterized by
nontrivial power laws:
\begin{equation}
P(t) \propto t^{-\delta},
\end{equation}
\begin{equation}
n(t) \propto t^{\eta},
\end{equation}
and
\begin{equation}
R^2(t) \propto t^z .
\end{equation}

I studied spreading in simulations beginning
with a localized density (typically, $\rho (i,0)
= \rho_{min} $ over the 10 - 20 sites nearest the origin,
and $\rho(i,0) = 0$ elsewhere), for
$D = b = 1$.  Time increments of $ 10^{-3} $ and
$10^{-2}$ were employed, as $\Dt = 10^{-4}$
resulted in excessive run times.
The trials (on the order of $10^3$
at each $a$-value of interest), ran to a maximum
time of 4000 (1000 in the studies employing
$\Dt = 10^{-2}$), and were performed on lattices
of  500 - 900 sites, sufficiently large that the active
region did not reach the boundaries.

Using
the criterion of asymptotic power laws at the
critical point, I estimate $a_c$ = 0.7210(5) for
$\Dt = 10^{-3}$, and $a_c$ = 0.568(1) for $\Dt = 10^{-2}$.
A plausible explanation  for the increase in $a_c$, as
$\Dt $ is reduced,
is that as $\rho_{min}$ becomes
smaller, and the truncation of the noise less severe, fluctuations
to zero density occur more readily.
When plotted versus $\Dt^{-1/2} $, the data for $a_c$
suggest a
finite limiting value $a_c(0)  \approx 0.8$,
with $a_c(\Dt) \approx a_c(0) + const. \times \sqrt \Dt$.
The present very limited data
(three points, for $D=b=1$),
are of course
insufficient to permit a firm conclusion in this
regard.

Figs 4 and 5 show the
evolution of the survival probability, mean population,
and root-mean-square spread ($x_{rms} \equiv \sqrt {R^2(t)}$),
for $\Dt = 10^{-3}$ and $10^{-2}$,
respectively.
In the latter case data from slightly off-critical
values are also plotted, which show some curvature.
{}From least-squares fits to the asymptotic
linear region in these log-log plots, I derive the
exponent estimates listed in Table 1.
Uncertainties, given by the figures in parentheses,
are subjective estimates based on the spread of
exponent values found in simulations with $a \approx a_c$.
While not of
high precision, the exponents found here compare rather
well against the known DP values.
Derivation of truly precise results will require longer runs
and larger samples, so that $a_c$ can be fixed more reliably,
and short-time corrections to scaling can be eliminated,
by means of a local-slope analysis.

Fig 6 shows the evolution of the mean density profile
in the critical system ($\Dt = 10^{-3}$, $a = 0.721$).  Following an
early build-up in the central region, the
profile broadens and becomes more sparse.
An interesting aspect of the late-time profile, which merits
further study, is its large spread, compared to a Gaussian
distribution.  (That is, $\overline{x^4} > 3\overline{x^2}$.)

\section{Discussion}

We have seen that some care is required in integrating
a field theory with multiplicative noise and an absorbing state.
To avoid negative densities and complete dominance of noise,
the SPDE must be regularized in some fashion.  The present
work shows discretization of
the field variable to be a suitable method for tempering
the equation.  Similar conclusions apply to the
associated SDE.  In fact, the method devised here yields an
accurate relaxation time for the latter problem.

Despite discretization of the density,
the present approach retains the
essential features of a continuum description.
The density approximates a continuous variable,
and spatial coupling accurs solely through diffusion.
Moreover, creation and annihilation are here
expressed in a naive mean-field-like manner (they are
represented, that is, by terms $\propto \rho$
and $\rho^2$).
Nontrivial critical behavior arises by virtue of properly scaled,
multiplicative noise.
The exponent values
derived from the discretized SPDE are in rather good
agreement with accepted values for DP in 1+1 dimensions,
arguing for the reliability of the method.
The numerical scheme proposed in this work may
therefore be of value in testing candidate theories for models
which have so far resisted analysis in continuum representation.
\vspace{2em}

\newpage

\noindent {\bf \large Acknowledgements}

I thank Michel Droz, Laurent Frachebourg, and John Cardy for
helpful discussions.  This work was initiated at the
Department of Theoretical Physics of the University of
Geneva, and continued, in part,
whilst visiting the
Isaac Newton Institute for Mathematical Sciences.
I am grateful for the warm hospitality I enjoyed at
these institutions.
Simulations employed the facilities of
the University Computing Center, City University of New York,
and of the University of Geneva.

\newpage
\begin{table}
\begin{center}
\noindent Table I.  Critical exponents from numerical integration
of the SPDE.
\vspace{2em}

\begin{tabular}{|c|c|c|} \hline
$\delta $ & $ \eta $ & $z $  \\
 \hline
\multicolumn{3}{|c|}{directed percolation} \\ \hline
0.1597(3)$^a$ & 0.317(2)$^b$ & 1.272(7)$^b$   \\ \hline
\multicolumn{3}{|c|}{present work:  $\Dt = 10^{-3}$} \\ \hline
0.15(1) & 0.28(1) & 1.18(2)  \\ \hline
\multicolumn{3}{|c|}{present work:  $\Dt = 10^{-2}$} \\ \hline
0.159(6) & 0.326(10) & 1.23(2) \\ \hline
\end{tabular}
\vspace{1em}

$^a$Ref.\cite{tdpjsp}   $\;\;\;\;^b$Ref. \cite{brower}
$\;\;\;\;\;\;\;\;\;\;\;$
\end{center}
\end{table}

\vspace{30em}
{}.

\newpage

\noindent{\bf Figure Captions}

\vspace{1em}
\noindent Fig. 1.  Evolution of the mean density in the discretized
SDE, Eq (\ref{onedeg}), for $a = 1.5$, $b=1$.   Squares:
$\Dt = 10^{-3}$; $+$: $\Dt = 2 \times 10^{-4}$; $\circ$: $\Dt = 10^{-4}$.
The inset shows a typical trial ($\Dt = 10^{-3}$).
\vspace{1em}

\noindent Fig. 2.  Steady state density {\em versus} $a$ in simulations of the
discretized SPDE, Eq (15), with $b=1$, and $\Dt = 10^{-4}$.
Solid squares: $D=1$; open squares, $D=10$.
\vspace{1em}

\noindent Fig. 3.  The data of Fig. 2 ($D=1$) plotted
{\em versus} $\Delta \equiv
a - a_c$, assuming $a_c = 0.769$.  The straight line,
fitted to the five points nearest $a_c$, has slope 0.295.
\vspace{1em}

\noindent Fig. 4.  Time-dependence of the suvival probability, $P$,
total density, $n$, and mean-square spread, $x_{rms}$, for
$a=a_c = 0.721$, $b=D=1$, and $\Dt=10^{-3}$.  The straight lines
are least-squares fits (for slopes see Table I).
\vspace{1em}

\noindent Fig. 5. Same as Fig. 4, but for $\Dt=10^{-2}$.
Dots: $a = 0.565$; open squares: $a=0.568$; circles:
$a= 0.570$.
\vspace{1em}

\noindent Fig. 6.  Average density profiles
for $D=B=1$ and $a = a_c = 0.721$, $\Dt = 10^{-3}$.
{}From narrowest to most broad: $t$ = 0, 500, 1000,  4000.

\end{document}